\documentclass[]{aa}

\usepackage{url}
\usepackage[breaklinks,colorlinks,citecolor=blue]{hyperref}
\usepackage{txfonts}
\begin{document}

   \title{The double Blue Straggler sequence in NGC2173: a field contamination artefact}


   \author{E. Dalessandro\inst{1}
   \and
   F.R. Ferraro\inst{2,1}
   \and
   N. Bastian\inst{3}
   \and
   M. Cadelano\inst{2,1}
   \and
   B. Lanzoni\inst{2,1}
   \and
  S. Raso\inst{2,1}
}

   \institute{INAF -- Astrophysics and Space Science Observatory Bologna, Via Gobetti 93/3 I-40129 Bologna, Italy\\
   \email{emanuele.dalessandro@oabo.inaf.it}
   \and
Dipartimento di Fisica e Astronomia, Via Gobetti 93/2 I-40129 Bologna, Italy
  \and
   Astrophysics Research Institute, Liverpool John Moores University, 146 Brownlow Hill, Liverpool, L3 5RF, UK
 }

   \date{Received August XX, 2018; accepted XXX XX, 2018}

 
  \abstract{Here we discuss the case of the double Blue Straggler Star (BSS) sequence recently detected in the young 
stellar cluster  NGC~2173 in the Large Magellanic Cloud by Li et al (2018). 
In order to investigate this feature we   
made use of two HST sets of observations, one (the same one used by Li et al.) probing the cluster central 
regions, and the other sampling the surrounding field. 
We demonstrate that when field star decontamination is applied, $\sim40\%$ of BSS population 
selected by Li et al. turns out to be composed by field stars interlopers. 
This contamination mainly affects one of the two sequences, which therefore disappears in the 
decontaminated colour-magnitude diagram. We analyse the result of 
tens different decontamination realisations: in none of them we find evidence of a double BSS sequence. 
Hence we conclude that NGC~2173 harbours a normal single (poorly populated) BSS sequence and that particular 
care needs to be devoted to the field decontamination process in any study aimed at probing 
stellar population features or star counts in the LMC clusters.    
}


\titlerunning{BSS in NGC2173}
\authorrunning{Dalessandro et al.}
   \maketitle



\section{Introduction}
Blue Straggler Stars (BSSs) represent the most numerous exotic population (i.e. not explainable 
in terms of passive evolution of single stars) in stellar systems. They are observed in globular (GCs; see for example
\citealt{ferraro03,ferraro12,ferraro18,dalessandro08a}) and open clusters \citep{mathieu09} as well as in dwarf galaxies \citep{monelli12}.
BSSs appear hotter and brighter than Turn-Off (TO) stars in optical colour-magnitude diagrams (CMDs) 
of stellar systems \citep{ferraro92,ferraro93,ferraro97},
thus mimicking a sparse sequence of younger and more massive objects. Observational evidence (see for example 
\citealt{shara97,ferraro06a,fiorentino14,brogaard18}) showed that indeed BSSs are up to twice more massive than TO stars in their
host clusters. 

BSSs can either be the result of mass accretion 
between two stars in a binary system \citep{mccrea64,zinn76} or the end products of stellar mergers induced
by collisions between single stars or binary systems \citep{hills76}. 
While the first process is common to any stellar environment being the result of the long-term evolution of binary systems, the second requires high density environments. As a consequence, in GCs, where the stellar density varies significantly from the center to the external regions, BSSs
can be generated by both processes with relative efficiencies depending on the local 
density \citep{fusipecci92,ferraro99}. In the high crowded cores of stellar clusters, both formation channels are expected 
to be active, although the mass-transfer process seems to be the most efficient one (see for example \citealt{knigge09}).

Interestingly, significant support to the co-existence of both collisional and mass-transfer BSSs in the same system, 
comes also from the detection of a double BSS sequence separated in color and magnitude
in four post-core collapse Galactic GCs, namely M~30 \citep{ferraro09}, NGC~362 \citep{dalessandro13} 
and M~15 (G. Beccari et al. 2018, in preparation), and NGC~1261,
which has been suggested to be in a post-core collapse bounce state \citep{simunovic14}.
The fact that the two sequences are well separated in the CMDs suggests that they were generated by a recent and short-lived event instead 
of a continuous formation process.

\citet{ferraro09} proposed that the origin of the double sequence can be related to 
the core-collapse (CC) process that is expected to largely enhance the probability of 
collisions over a relatively short period of time (of the order of few tens of Myr), thus promoting the formation of collisional BSSs. 
In fact, during CC the central density rapidly increases, thus producing an increase in gravitational interactions 
\citep{meylan97} able to trigger the formation of new BSSs through both direct stellar collisions and mass 
transfer activity induced by dynamical encounters involving primordial binaries 
(e.g., \citealt{leonard89,hurley05,banerjee16}).
Thus, the CC process could in principle generate a nearly coeval population of BSSs with different masses 
that in the CMD appears as a  well-defined, tight sequence in addition to the ``normal'' (spread) 
BSS population constantly generated by the mass transfer process in primordial binaries, 
which are present in any environment.  
Indeed, the location of the collisional sequences in the CMD can be well reproduced 
by collisional models \citep{sills09}. Analogously to what is commonly done in the Main Sequence 
fitting dating method, this allows us to use the extension in luminosity of the collisional sequence 
to date the epoch of the formation of the collisional BSSs, thus providing an hint to the epoch of 
CC in the parent cluster (\citealt{ferraro09}). 
Following these arguments \citet{ferraro09} and \citet{dalessandro13} concluded that the CC event 
occurred recently (1-2 Gyr ago) in both M~30 and NGC~362.
The red BSS sequence is instead compatible with
models of binary systems undergoing mass-transfer (\citealt{tian06,xin15}, but see the discussion in \citealt{jiang+17}).    
 
Recently, \citet{li18} have claimed that also the young ($t\sim2$ Gyr) 
GC NGC~2173 in the Large Magellanic Cloud (LMC) shows a double sequence of BSSs.
NGC~2173 is the only non post-core-collapse cluster suggested to show a double BSS sequence so far. This result can have therefore important implications
in our understanding of BSS double sequence formation.
In fact, because of the cluster's low central density and age, the authors argued that collisions are not a
viable scenario to explain the blue BSS sequence in this case. On the contrary, they suggest 
that most likely the two sequences are populated by binaries with different mass-ratios. However, 
they admit that it is unclear how this scenario can reproduce the observed discrete color 
and magnitude distribution.

The significance of the observational results presented by \citet{li18} is strongly hampered by the fact 
that they do not take into account field star contamination.
In LMC clusters, field interlopers can potentially have a large impact on the BSS region in the CMD, because of the
large young LMC stellar component.
In this work we re-analyse the BSS population properties and distribution taking into account the effect of field star
contamination. To this aim, we used two different HST data-sets, sampling both the cluster central regions 
and the surrounding LMC field. We find that about half of the BSS total sample used by \citet{li18}
is actually populated by likely non-member stars. 
More interestingly, we show that the decontamination analysis essentially remove all the BSS along the 
red sequence, thus demonstrating that the double sequence claimed by \citet{li18} is essentially an 
artefact due to the contamination of the LMC field stars.

\begin{figure}
\includegraphics[width=9cm]{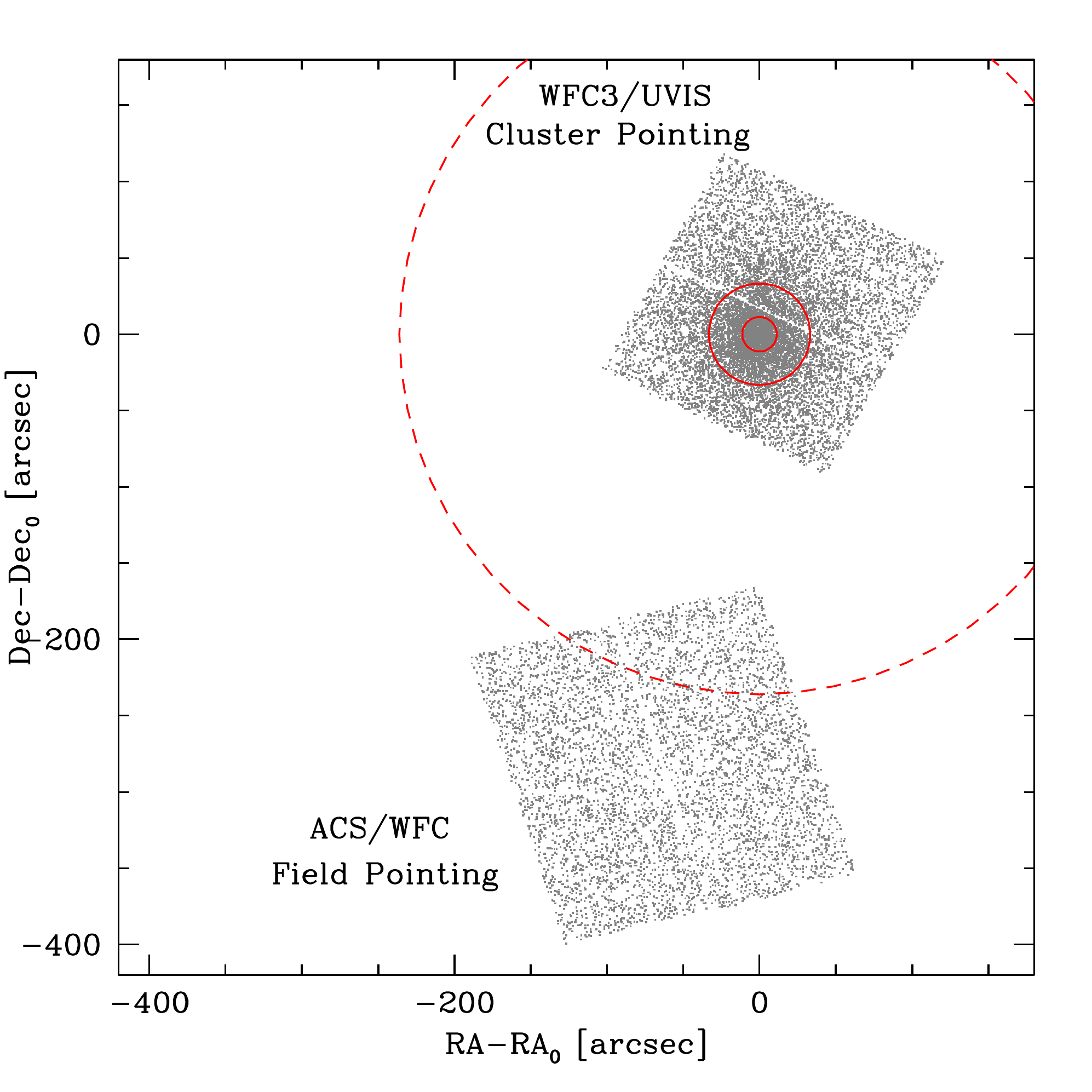}
\caption{Map of the WFC3/UVIS and ACS/WFC fields of view. The cluster core, half-light
and tidal radii (see Appendix) are shown in red. }
\label{fig:map}
\end{figure} 

\section{Observations and data analysis}
We used a combination of Hubble Space Telescope Wide Field Camera 3 - UVIS channel (WFC3/UVIS) and
Advanced Camera for Survey - WFC (ACS/WFC) images obtained through proposal GO12257 (PI: Girardi).
WFC3/UVIS data represent the primary observations and sample the cluster innermost regions 
({\it Cluster Pointing}), while ACS/WFC data have been
obtained as parallel observations ({\it Field Pointing}) and sample a region located at $\sim300\arcsec$ southward
from the cluster (see Figure~\ref{fig:map}).

\begin{table*}
\begin{center}
\caption{Summary of the HST data-set used in this work.}
\begin{tabular}{ c  c  c  c }
\hline
\hline
Instrument & Filter & ${\rm t}_{{\rm exp}}$ (s) & Camera\\
\hline
{\small Cluster Pointing}      & {\scriptsize F336W} &  { \scriptsize $1 \times 800 + 2\times 700$}    & { {\small WFC3/UVIS}}\\
{}			       & {\scriptsize F475W} &  { \scriptsize $1 \times 120 + 2\times 700$}    & {}\\
{}			       & {\scriptsize F814W} &  { \scriptsize $1 \times 20 +  3\times 700$}    & {}\\

\hline
{ \small Field Pointing}	  & { \scriptsize F475W}  & { \scriptsize $2 \times 500 + 1 \times 700$}  & {{\small ACS/WFC}}\\ 
{}			    & { \scriptsize F814W}  & { \scriptsize $1 \times 20 + 1 \times 600 + 1\times 690+ 2\times 700$}	  & {}\\
\hline

\hline
\hline
\end{tabular}
\\
\label{dataset}
\end{center}
\end{table*}

The most relevant details about the adopted sets of images are summarised in Table~1.
For both data-sets, an appropriate dither pattern of a few arcseconds has been adopted for each pointing 
in order to fill the inter-chip gaps and avoid spurious effects due to bad pixels.

\begin{figure}
\includegraphics[width=9cm]{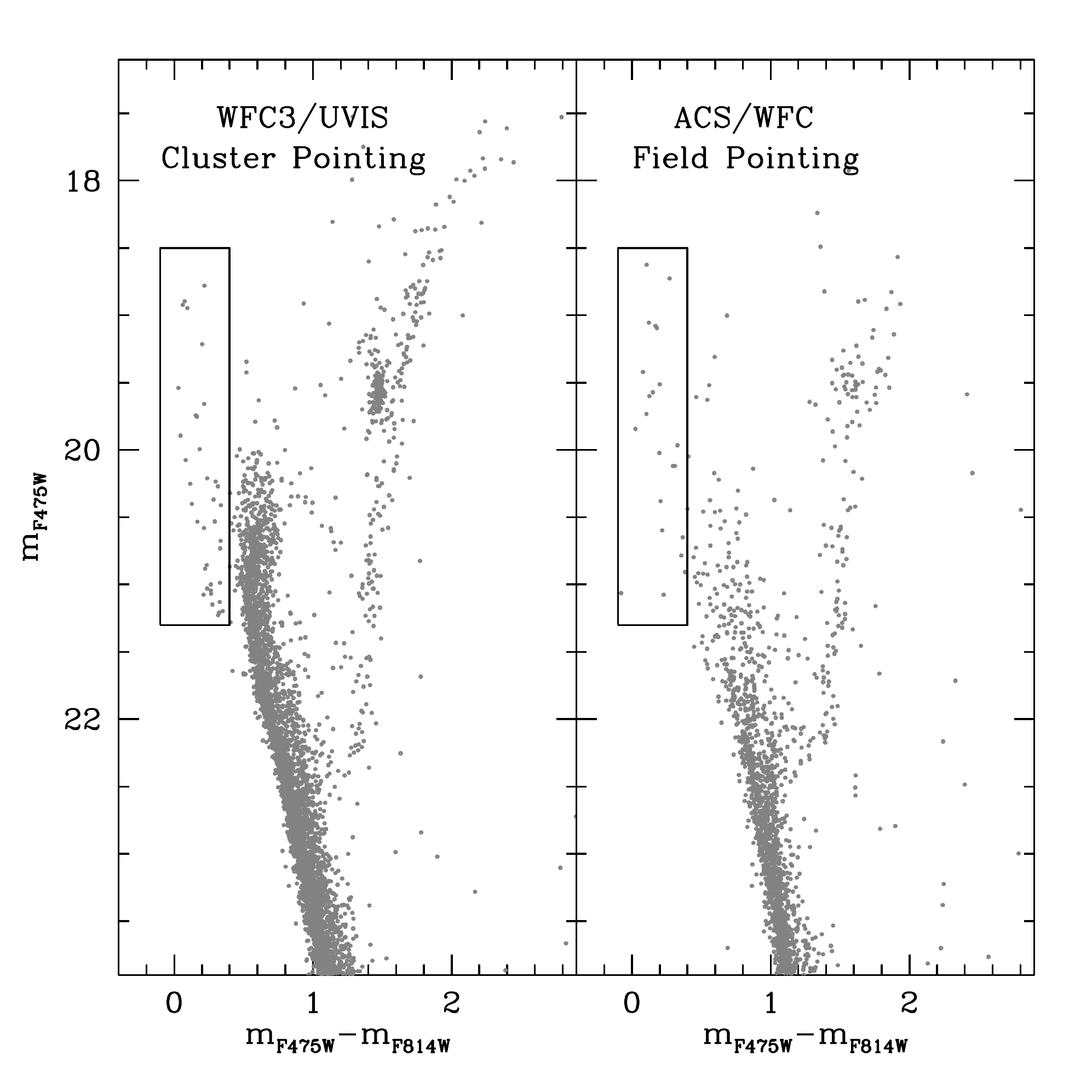}
\caption{($m_{F475W},m_{F475W}-m_{F814W})$ CMDs of the {\it Cluster Pointing} and {\it Field Pointing}. 
The black boxes highlight the BSS region that has been used in the analysis (see Section~3). }
\label{fig:opt-cmds}
\end{figure} 

The photometric analysis was performed on the {\tt FLC} images (which are corrected for charge transfer efficiency) 
{\tt pixel area
map} corrected images by using {\tt DAOPHOT IV} \citep{stetson87} and relative routines, following 
the approach used in other works of our group
(see \citealt{dalessandro18a,dalessandro18b} for a recent reference). 
Briefly, point spread function (PSF) models were derived for each image and chip by using some tens of stars and they were applied to stars whose flux peaks are above $3\sigma$ from the local background.
A master list including stars detected in at least four images was then created. 
At the corresponding positions of stars in the master-list, a fit was forced with 
{\tt DAOPHOT/ALLFRAME} \citep{stetson94} in each frame of the two data-sets. 
For each star thus recovered, multiple magnitude estimates obtained in each chip were homogenised by using 
{\tt DAOMATCH} and {\tt DAOMASTER}, and their weighted mean and standard deviation were finally adopted as 
star magnitude and photometric error.

Instrumental magnitudes of both the WFC3 and ACS catalogs 
were calibrated onto the VEGAMAG photometric system by using the recipes and zero-points reported in the HST web-sites.
Instrumental coordinates were first corrected 
for geometric distortions by using equations 
by \citet{bellini09} for the WFC3/UVIS data and the most updates Distortion Correction Tables 
({\tt IDCTAB}) provided on the dedicated page of the Space Telescope Science Institute for the ACS/WFC images.
Then they were reported to the absolute coordinate system ($\alpha, \delta$) as defined by the World Coordinate System
of the HST images.
The final color magnitude diagrams for both data-sets are shown in Figure~\ref{fig:opt-cmds}.
We note that, in agreement with \citet{li18}, a double BSS sequence can be easily distinguished in the {\it Cluster Pointing} CMD.

\section{Density profile and structural parameters}
In order to derive cluster structural parameters 
we built the cluster number density profile by using both the {\it Cluster} and the {\it Field} HST data-sets.
These information, and the extension of the cluster in particular, are key for the decontamination procedure described in
Section~4.

\begin{figure}
\includegraphics[width=9cm]{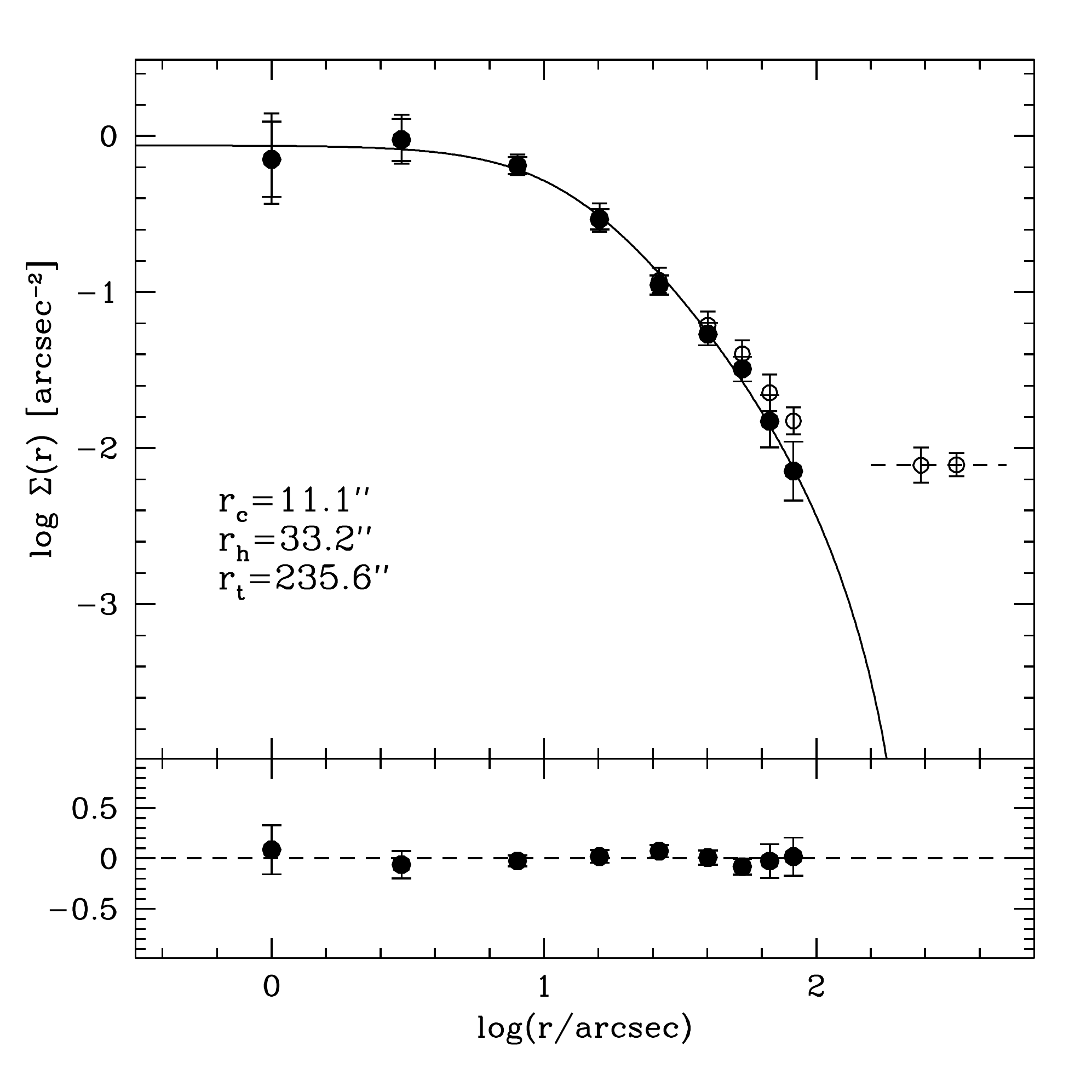}
\caption{Observed star count density profile of NGC2173 (open circles). 
The dashed line represents
the density value of the background as derived by averaging the two outermost radial bins. The black filled dots are densities obtained after background subtraction. The best-fit single-mass King model is overplotted to the observations (black solid line). The lower panel shows the residuals between the observations and the best-fit model.}
\label{fig:dens}
\end{figure}

The analysis has been performed following the procedure fully described in Miocchi et al. (2013; see also \citealt{lanzoni10}). We used as center of the cluster the one obtained by \citet{li18}.
We considered 11 concentric annuli centred on the cluster center, each one divided into a number of sub-sectors ranging from two to four. 
In each sub-sector, 
we counted the number of stars with $m_{F606W}<21$. 
The projected stellar density in each annulus 
is the mean of the values measured in each sub-sector and the uncertainty has been estimated from the variance 
among the sub-sectors.  
The observed density profile is shown in Figure~\ref{fig:dens} (open circles). It smoothly declines as a function of distance
and then it flattens at distance $d>200\arcsec$ due to the contribution
of the LMC field background. This has been estimated by averaging the two outermost values, 
and it has been subtracted to the observed distribution to obtain the cluster (decontaminated) density 
profile (filled circles). 
 
We then derived the cluster structural parameters 
by fitting the observed density profile with a spherical, isotropic, single-mass \citet{king66} model. The best-fit model 
results in a cluster with a King dimensionless potential $W_0 = 6.1$, corresponding to a concentration parameter $c=1.28$
a core radius $r_c = (11.1\pm1.2)\arcsec$, a half-mass radius $r_h=(33.2\pm3.3)\arcsec$, 
a tidal radius $r_t=(235.6\pm23.5)\arcsec$. Assuming a distance to the LMC of 50 kpc \citep{pietr13}, these values correspond to 2.7pc, 8.1pc and 57.1pc respectively.
While the derived values of $r_c$ and $r_h$ are in quite good agreement with literature estimates, $r_t$ results to be significantly smaller than that obtained by \citet{vdm05} and \citet{li18}, who quote 
95.5pc and 131.30pc, respectively.
Such a discrepancy is likely due to the lack of an appropriate control area allowing to properly 
sample and subtract the LMC background.

\begin{figure}
\includegraphics[width=9cm]{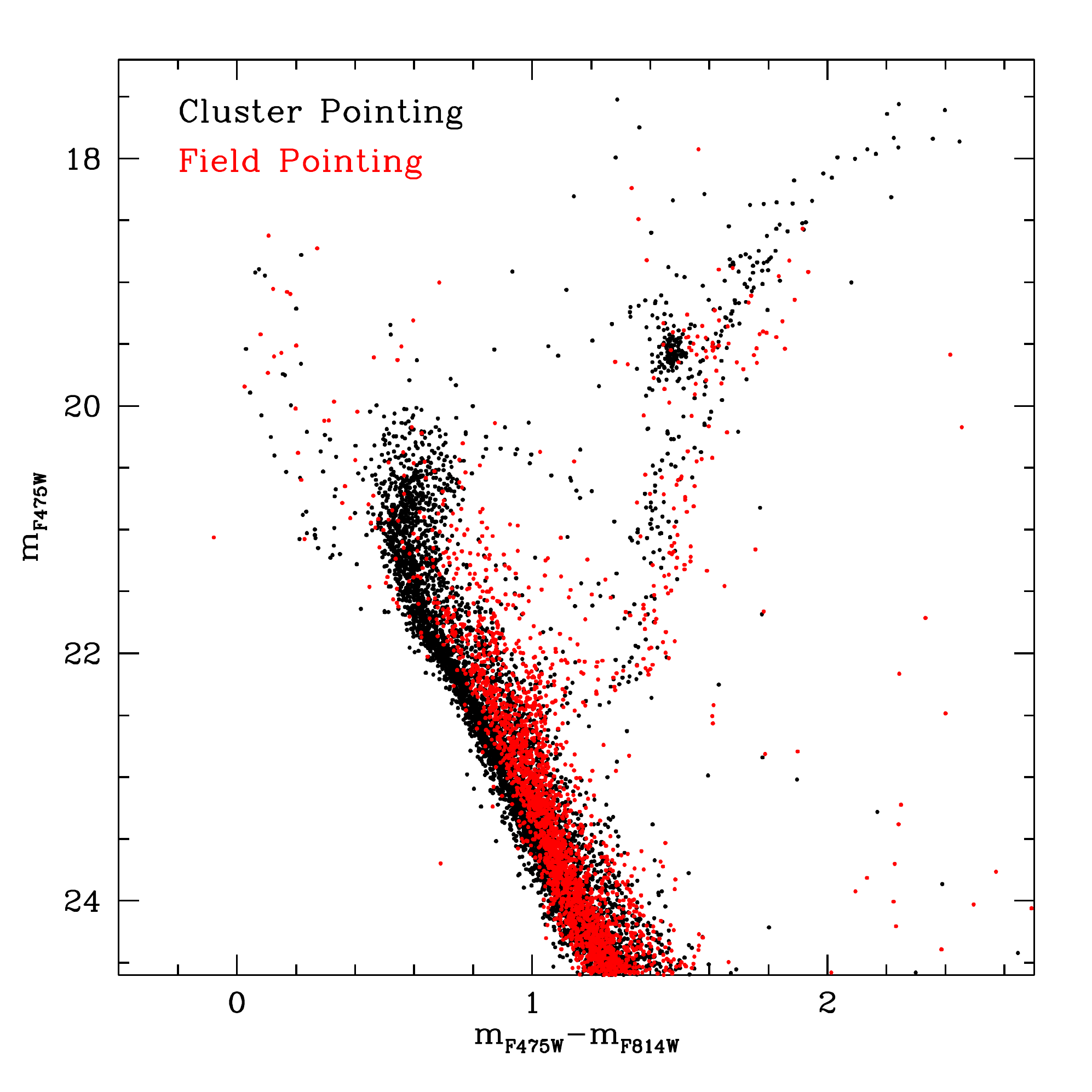}
\caption{($m_{F475W},m_{F475W}-m_{F814W})$ CMD of the {\it Field Pointing} (in red) superimposed to the 
{\it Cluster Pointing} CMD (in black) for a direct comparison. Clearly, the SGB/RGB at $m_{F475W}-m_{F814W}>0.8$ and $m_{F475W}>21$,  
as well as the BSS region are significantly contaminated by LMC stars.}
\label{fig:comp-cmds}
\end{figure}

\section{Color magnitude diagrams decontamination}
The CMDs of the vast majority of stellar clusters in the Magellanic Clouds are strongly contaminated 
by field star interlopers. In fact, as shown in Figure~\ref{fig:opt-cmds}
for the case of NGC~2173, cluster evolutionary sequences are overlaid to similarly populated LMC 
star sequences. As a consequence, a proper decontamination is key  when studies on specific portions
and features in clusters' CMDs are performed and/or when stellar population number counts are concerned. 

Unfortunately, given the distance of the Magellanic Clouds a detailed separation between field and cluster stars based on 
proper motion is possible only for a few cases. Moreover accurate Gaia DR2 proper motions 
data are available only for the brightest stars. 
As a consequence, to assess the impact of field contamination on the CMD of NGC~2173 we used a  statistical approach based 
on the comparison between the distribution of stars in the cluster CMD, at different radial distances from the cluster center,
and that of a region representative of the surrounding LMC field. 
As shown in Figure~\ref{fig:map}, the available ACS/WFC {\it Field Pointing} is ideal to perform a suitable decontamination 
of the program cluster. In fact, the ACS/WFC FOV provides a relatively large area (which is $\sim50\%$ larger 
than the WFC3/UVIS 
{\it Cluster Pointing} FOV) to sample the LMC star distribution. Moreover, it is located beyond the cluster tidal 
radius (see Section~3) thus ensuring that any contribution from cluster stars (if any) is negligible.

\begin{figure}
\includegraphics[width=9cm]{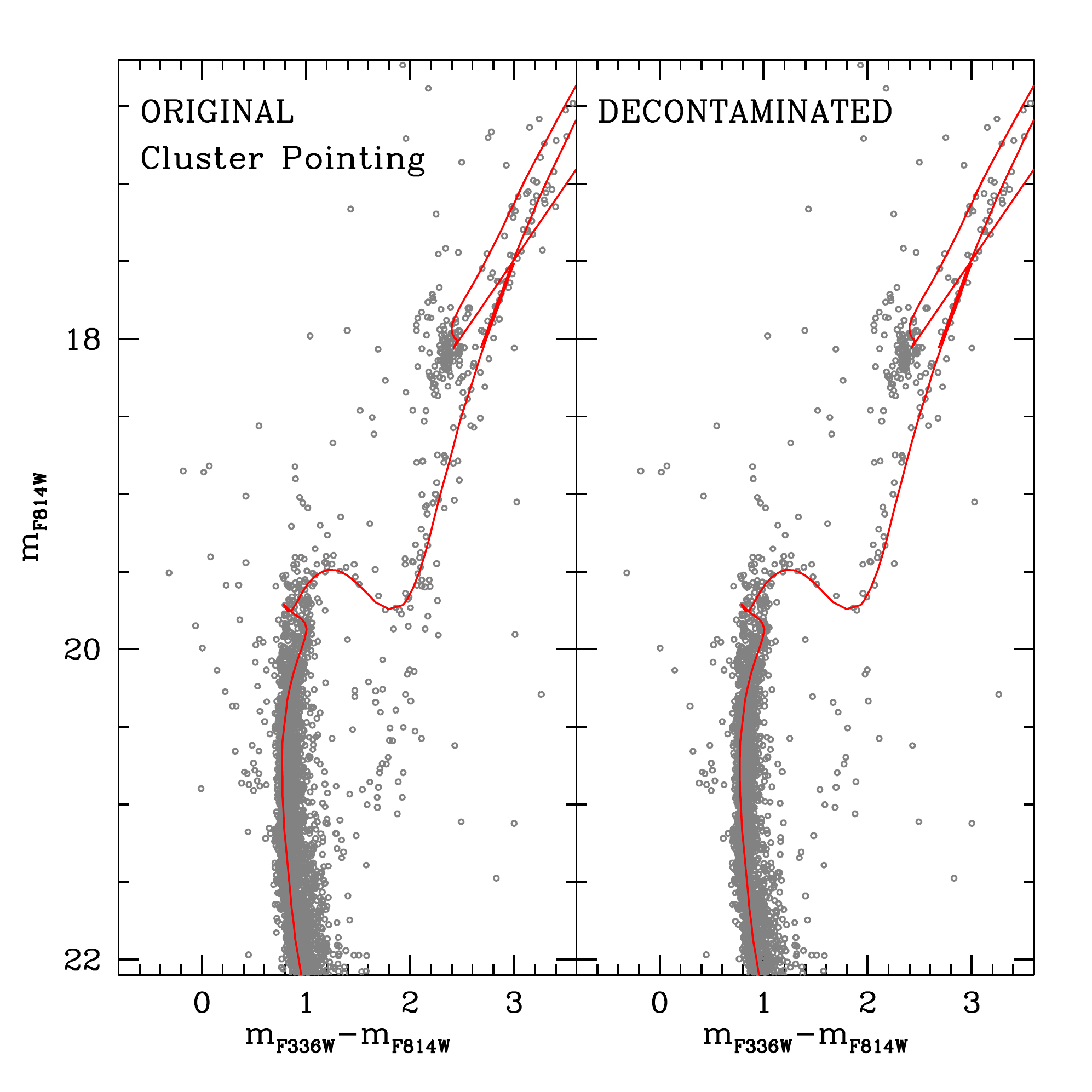}
\caption{Comparison between the observed {\it Cluster Pointing} CMD and one resulting from the statistical decontamination procedure
described in Section~3.}
\label{fig:deco}
\end{figure}

\begin{figure*}
\centering
\includegraphics[width=14cm]{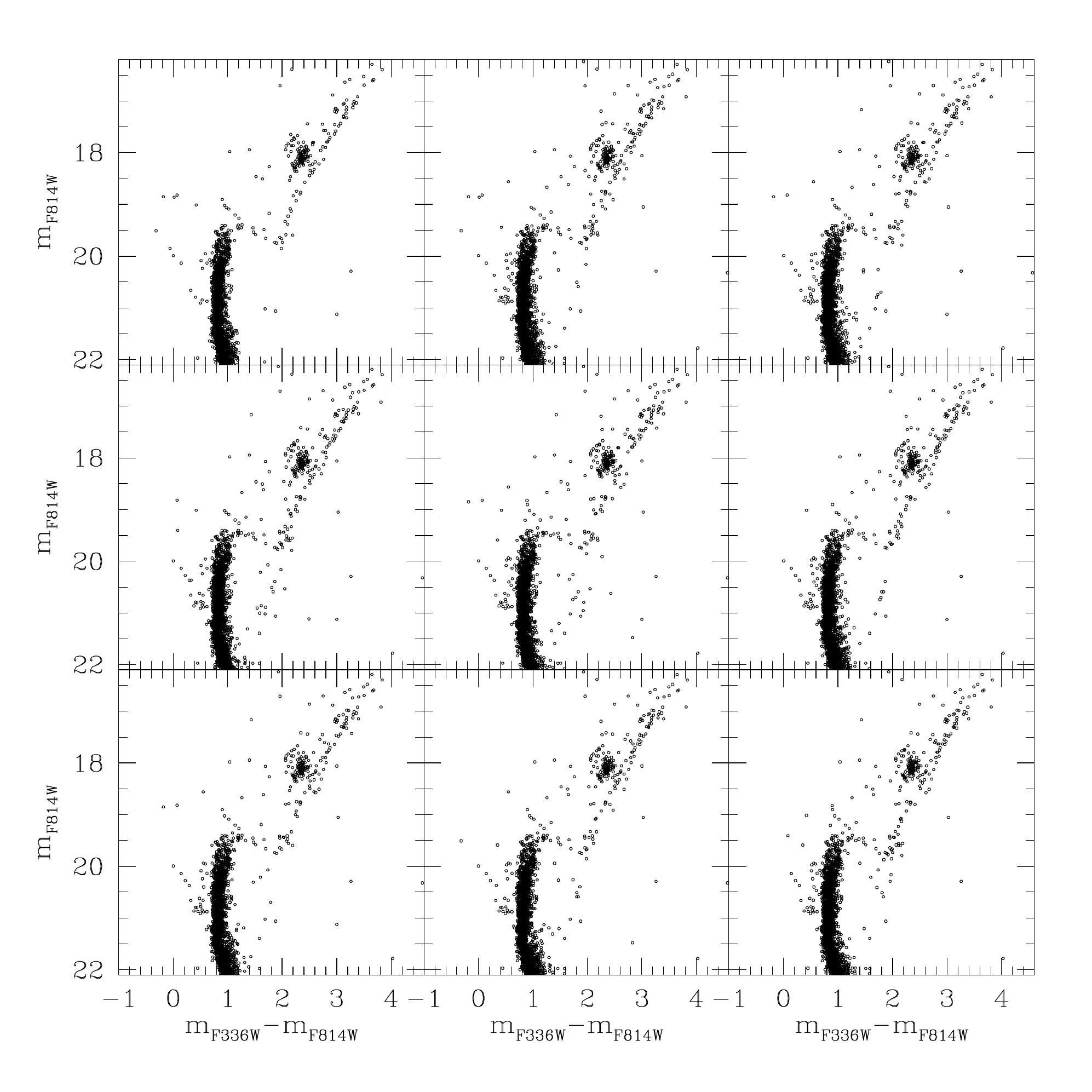}
\caption{Nine decontaminated CMDs obtained as described in Section~3, using different grid's cell widths and limits. 
In all cases there is no significant evidence of a double BSS sequence.}
\label{fig:multi}
\end{figure*}

A first clue about the level of the LMC field contamination affecting the {\it Cluster} CMD can be obtained by 
applying the method described in \citet[][see also \citealt{ziri16} for a similar application]{knoetig14}
which allows us 
to calculate the probability of a star in the {\it Cluster Pointing} CMD to belong to the field population.
We divided both the {\it Cluster} and the {\it Field} CMDs in a regular grid of $0.3\times0.25$~mag$^2$ 
and we counted the number of stars falling within each cell. 
The number counts of likely field stars were then corrected for the different 
area covered by the two data-sets. We then used equations 23 in \citet{knoetig14} 
to estimate the probability that stars observed in a given cell are cluster members.
We refer the reader to that paper for all relevant details about the method.

This approach has the advantage to be applicable to cells with small and large numbers.  
The analysis shows that there are at least two regions in the {\it Cluster} CMD significantly contaminated by foreground LMC stars: 
the one including stars with ($m_{F475W}-m_{F814W})>0.8$ and $m_{F475W}>21$, where at least two LMC
sub-populations older than NGC~2173 can be distinguished, and the BSS region, in which 
the member probability is below $20\%$.

In Figure~\ref{fig:opt-cmds} we highlight with a rectangular box the BSS region
in the ($m_{F475W},m_{F475W}-m_{F814W}$) CMDs of both the {\it Cluster} and the {\it Field} pointings.
Even from a preliminary visual inspection it is evident that a significant contribution of the LMC field 
is expected in that region. In fact, we count 37 stars ($\pm6.2$ as Poissonian error) in the WFC3 {\it Cluster Pointing} 
and 21 in the ACS/WFC {\it Field Pointing}. After rescaling this latter value to the WFC3/UVIS area, 
we obtain a total number of contaminants of $15\pm3.9$, which yields a possible number of likely BSS cluster members
of $22\pm7.2$ in the {\it Cluster Pointing} (corresponding to roughly $60\%$ of the observed sample).

In order to allow a direct comparison between the cluster and the field sequences, in Figure~\ref{fig:comp-cmds},
we co-added the two CMDs, highlighting in red the stars measured in the {\it Field Pointing}. 
The matching of the sequences is impressive
and clearly indicates the regions of the CMD which are mainly affected by the LMC field contamination,
fully confirming the analysis discussed above. 
In fact, it is evident that the two SGB/RGB visible at  ($m_{F475W}-m_{F814W})>0.8$ and $m_{F475W}>21$ 
are essentially due to the LMC field. Moreover, both the Red Giant Branch and the Helium Red Clump of the cluster 
are affected by field contamination, as well as the BSS region. 
In particular, from the comparison it also emerges that the BSSs in the {\it Field} and the {\it Cluster} CMDs are 
distributed in a different way. 
In fact, field candidate BSSs (or young populations) 
tend to be preferentially distributed in the brightest and 
reddest portion of the box, thus suggesting that red BSSs in the {\it Cluster} CMD are less likely to be cluster  
members than the blue ones.

These simple arguments demonstrate that a very detailed analysis is needed to 
assess the possible presence of a double sequence, as that claimed by \citet{li18}.

To this aim we have statistically decontaminated the ($m_{F475W},m_{F475W}-m_{F814W}$) CMD by using the following approach.
We split the {\it Cluster Pointing} in five concentric annuli centred on the cluster center.
We then partitioned the CMD corresponding to each annulus in the same grid of cells discussed above for the {\it Field Pointing} CMD.
The same number of stars counted in the corresponding 
cell of the {\it Field} CMD was then randomly removed from the {\it Cluster} cell, 
accounting for the difference in the size between the area covered by ACS and the WFC3 pointings.
The stars surviving this analysis correspond to the likely cluster members and constitute the decontaminated CMD.
The original CMD and the one resulting from the decontamination analysis are shown in Figure~\ref{fig:deco} 
(to ease the comparison with \citealt{li18} we show the results in the ($m_{F336W},m_{F336W}-m_{F814W}$) CMD). 
The analysis has efficiently removed highly probable LMC stars and thus the evolutionary sequences appear to be much 
better defined in the decontaminated CMD. 
As expected, from the arguments discussed above,
only about 25 BSSs survived the decontamination analysis. 
More importantly, the remaining BSS population is quite scarce and shows no evidence of a double sequence.

We performed the above analysis several tens of times by changing the cells' dimensions and the grids' limits.
Some of the resulting decontaminated CMDs are shown in Figure~\ref{fig:multi} as examples.
As expected, while the exact number and position of the surviving BSSs may differ from 
one realisation to the other, the two main results remain:
i) the total fraction of BSS cluster members in the {\it Cluster Field} is only about $60\%$ of the observed one; 
ii) in none of the realisations we find evidence of a double BSS sequence.

\section{Summary and Conclusions}
We used observations available in the HST Archive
to perform a detailed decontamination of the CMD of the young cluster NGC~2173 in the LMC, with the specific aim of 
assessing the existence of the double BSS sequence as recently claimed by \citet{li18}. Our analysis demonstrates
that the detected feature is an artefact due to the contamination by LMC field stars. 
NGC~2173 turns out to be a young cluster with a single and poorly populated BSS sequence, which can likely be the result of binary evolution.
As a consequence, the case of NGC~2173 is not relevant for the understanding and the discussion on the origin of the double 
BSS sequences observed so far in a few old globular clusters (\citealt{ferraro09,dalessandro13,simunovic14}, Beccari et al. 2018). 

We used the decontaminated CMD shown in the right panel of Figure~\ref{fig:deco} to select 
likely member BSSs and reference populations. 
We find that the BSS radial distribution in NGC~2173 is consistent (within the uncertainties related to the 
small number of stars and the decontamination process) with that of normal Turn Off cluster stars. 
This is not surprising since the BSS originated by the evolution of primordial binaries are expected to have, 
at their origin, a radial distribution indistinguishable from other lighter stars. 
Such a radial distribution can be maintained for a long time in dynamically unevolved clusters, 
where the time-scale of the BSS sedimentation toward the cluster center due to dynamical friction is particularly long. 
This is indeed observed in many dynamically young Galactic clusters (see the cases of Omega Centauri \citealt{ferraro06b}, 
NGC~2419 and NGC~6101 \citealt{dalessandro08b,dalessandro15}) even after several Gyr from their formation 
(see the discussion in \citealt{ferraro12,ferraro18,lanzoni16}). Hence
no peculiar phenomena need to be advocated to explain the BSS population of NGC~2173. 
Instead this case clearly points out the importance and the necessity of an appropriate field 
decontamination to properly assess any feature or star population counts in heavily contaminated stellar 
aggregates, as the stellar clusters in the LMC.

\vspace{1.0cm}
ED acknowledges financial support from the Leverhulme Trust Visiting Professorship Programme VP2-2017-030.
ED is also grateful for the warm hospitality of LJMU where part of this work was performed.
NB gratefully acknowledges funding from the ERC under the European Union's Horizon 2020 research and innovation programme via 
the ERC Consolidator Grant Multi-Pop (grant agreement number 646928, PI Bastian). NB is a Royal Society University Research Fellow.
The authors thank the anonymous referee for his careful reading of the paper and useful comments.




\end{document}